# Fluorination Effects on the Structural Stability and Electronic Properties of sp$^3$ Type Silicon Nanotubes

*Alon Hever, Jonathan Bernstein, and Oded Hod*[*]

Department of Chemical Physics, School of Chemistry, the Sackler Faculty of Exact Sciences, Tel-Aviv University, Tel-Aviv 69978, Israel

**Abstract**

A density functional theory study of the structural and electronic properties and relative stability of fluorinated sp$^3$ silicon nanotubes and their corresponding silicon nanowires built along various crystallographic orientations is presented. The structural stability is found to increase linearly with the fluorine surface coverage and for coverages exceeding 25% the tubular structures are predicted to be more stable than their wire-like counterparts. The bandgaps of the fully fluorinated systems are lower than those of their fully hydrogenated counterparts by up to 0.79 eV for systems having a relatively low silicon molar fraction. As the silicon molar fraction increases these differences appear to reduce. For mixed fluorination and hydrogenation surface decoration schemes the bandgaps usually lie between the values of the fully hydrogenated and fully fluorinated systems. Furthermore, the bandgap values of the silicon nanotubes are found to be more sensitive to the fluorine surface coverage than those of the silicon nanowires. These results indicate that surface functionalization may be used to control the stability of narrow quasi-one-dimensional silicon nanostructures and opens the way towards chemical tailoring of their electronic properties.

**Introduction**

Due to their unique and diverse physical properties quasi-one-dimensional silicon nanostructures show great promise to serve as active components in various nanoscale devices including sensitive chemical and biological detectors,[1-7] electronic components such as nanoscale field-effect transistors,[8,9] optoelectronic devices,[10-12] solar cells[13-16], and water splitting photocatlysts.[17] Since different applications require different electronic properties it is desirable to identify efficient and accessible routes to control and tailor the electronic structure of such systems. To this end, experimental and theoretical studies of silicon nanowires (SiNWs) have revealed several factors that may influence the size and character of their bandgap. These include the diameter of the SiNW,[18-26] its growth orientation,[20,23,26,27] cross sectional shape,[28,29] surface reconstruction,[26] chemical passivation[27,30-33] and doping,[34] and strain effects.[35-37]

To gain control over the geometrical parameters of SiNWs several fabrication methods have been developed including laser ablation and electron beam evaporation metal-catalytic vapor-liquid-solid methods,[32,38-42] oxide-assisted catalyst-free approaches,[43-45] as well as solution based techniques.[46] These methods yield wires with different crystallographic orientations and dimensions scaling down to diameters which are in the single nanometer regime.[32,38,43,47] The effects of chemical modifications on the physical properties of SiNWs have been investigated experimentally in several recent studies, as well. Here, it was shown that hydrogen passivated SiNWs can be achieved via removal of the oxide layer often decorating their surface.[39,43,47-55] Alternatively, surface hydrogenation can be achieved by using hydrogen as a carrier gas in the chemical vapor deposition process.[42]

From the theoretical perspective, hydrogen passivation of SiNWs is attractive in terms of lowering the computational burden. Hence, many of the earlier theoretical studies of SiNWs have adopted hydrogen passivated SiNWs models.[20,24,26,56,57] These were followed by several recent studies using hydrogenated SiNWs models as reference systems for SiNWs with other chemical passivation schemes.[27,30,31,58-60] It was shown that decoration with hydroxyl and amine groups tends to cause diameter dependent bandgap red shifts with respect to the hydrogen passivated systems.[27,30]

The effects of surface halogenation of [110] SiNWs with diameters varying from 0.6-3 nm were studied by Leu *et* al. showing that increased halogen surface coverage tends to decrease the bandgap of the SiNW.[31] As may be expected, larger bandgap variations were found for narrower wires which have a larger surface to volume ratio. Furthermore, the identity of the halogen atom was found to impact the resulting bandgap of the system where larger halogen atoms resulted in more pronounced effect on the bandgap. Bondi et al.[53] have shown that for SiNWs grown along the [111] orientation, fluorine surface decoration results in increased band density. Studying the effects of crystallographic orientation, Ng *et* al.[61] have reported bandgap decrease of up to 1 eV with increasing [100], [110], [111], [112] SiNW surface coverage of fluorine and hydroxyl groups. For the fluorinated SiNWs the CBM and VBM energies were found to strongly decrease with respect to the fully hydrogenated counterparts.[⊥]

Going beyond SiNW structures, we have recently presented a computational study exploring the structural and electronic properties of fully hydrogenated $sp^3$-type silicon nanotubes (SiNTs) grown along various crystallographic directions.[70] Here, all SiNTs studied were found to be less (meta-)stable than the corresponding SiNWs while possessing an increased bandgap. In light of the recent intensive work on chemical decoration of SiNWs it would be interesting to explore the effects of surface functionalization on the structural stability and electronic properties of SiNTs. In the present work, we address this issue showing that surface fluorination is expected to dramatically increase the relative stability of quasi-one-dimensional silicon structures as compared to their hydrogenated counterparts. Furthermore, we show that surface

---

[⊥] It should be noted that halogenated silicon surfaces are often regarded as highly reactive and therefore should be treated under inert conditions. (62) Buriak, J. M. Organometallic Chemistry on Silicon and Germanium Surfaces. *Chemical Reviews* **2002**, *102*, 1271-1308. Specifically, plasma etching of silicon surfaces with molecular fluorine was shown to result in $SiF_2$ terminating groups whereas high exposure to atomic fluorine induce tetrafluorosilane desorption from the surface. (63) Mucha, J. A.; Donnelly, V. M.; Flamm, D. L.; Webb, L. M. Chemiluminescence and the reaction of molecular fluorine with silicon. *The Journal of Physical Chemistry* **1981**, *85*, 3529-3532, (64) Stinespring, C. D.; Freedman, A. Studies of atomic and molecular fluorine reactions on silicon surfaces. *applied physics letters* **1986**, *48*, 718-720, (65) Nakamura, M.; Takahagi, T.; Ishitani, A. Fluorine Termination of Silicon Surface by F2 and Succeeding Reaction with Water. *Jpn. J. Appl. Phys.* **1993**, *32*, 3125-3130, (66) Tu, Y.-Y.; Chuang, T. J.; Winters, H. F. Chemical sputtering of fluorinated silicon. *Physical Review B* **1981**, *23*, 823-835, (67) Wu, C. J.; Carter, E. A. Structures and adsorption energetics for chemisorbed fluorine atoms on Si(100)-2×1. Ibid.**1992**, *45*, 9065-9081, (68) Aliev, V. S.; Kruchinin, V. N.; Baklanov, M. R. Adsorption of molecular fluorine on the Si(100) surface: an ellipsometric study. *Surface Science* **1996**, *347*, 97-104, (69) McFeely, F.; Morar, J.; Himpsel, F. Soft x-ray photoemission study of the silicon-fluorine etching reaction. Ibid.**1986**, *165*, 277-287.

chemistry may be used as a simple means to control the electronic properties of SiNWs and SiNTs.

To this end, we reconsider the eight SiNWs and SiNTs models studied in Ref. [70] with growth orientations along the [100], [110], [111] and [112] silicon crystal directions (see Fig. 1). Using the hydrogenated systems as reference we then partially or fully substitute passivating hydrogen atoms with fluorine atoms. For the partial substitution schemes, we consider fluorine surface coverage of 25, 50, and 75% where the substitution sites are determined randomly and for each coverage three different random substitution schemes are considered. We note that the experimental realization of narrow SiNTs has been recently achieved[71-77] showing the ability to control their surface chemistry.[75,76]

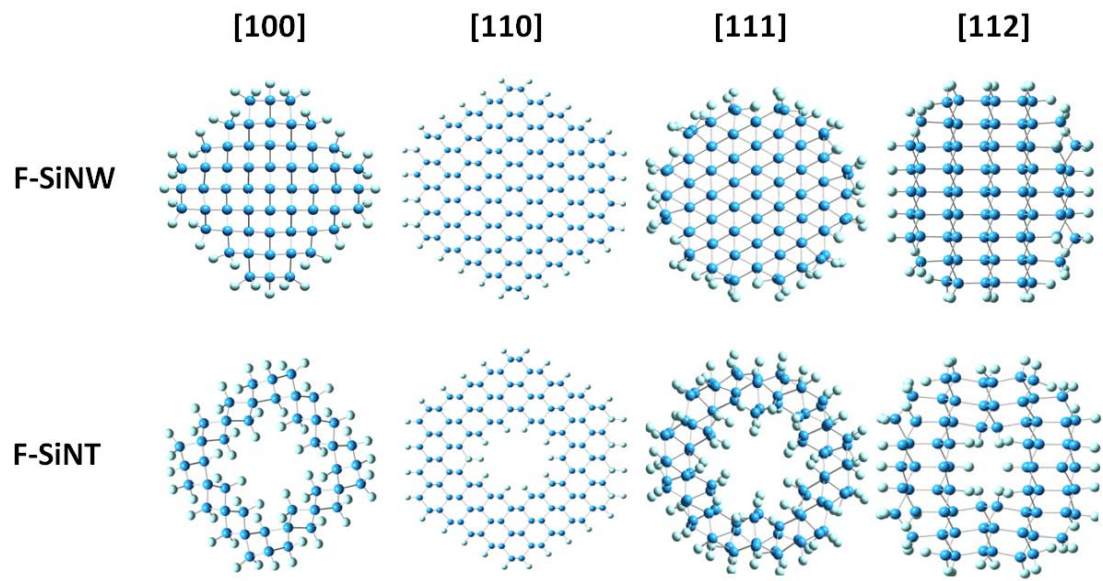

Figure 1: *Schematic representation of various fluorinated SiNWs and SiNTs carved out of bulk silicon along the [100], [110], [111] and [112] crystallographic orientations. Light-blue and gray spheres represent silicon and fluorine atoms, respectively.*

All calculations have been performed using density functional theory (DFT) as implemented in the *Gaussian* 09 suite of programs.[78] Three different functional approximations have been considered, namely, the local density approximation (LDA),[79,80] the PBE realization of the generalized gradient approximation,[81] and the screened exchange hybrid density functional of Heyd, Scuseria, and Ernzerhof (HSE).[82-85] The latter functional has been tested for a wide set of materials and was shown to accurately reproduce experimental structural parameters and bandgaps.[70,86,87] Initial geometry optimizations have been performed using the LDA with the 3-21G atomic centered Gaussian basis set. Further geometry relaxation has been performed for each functional approximation separately using the double-$\zeta$ polarized 6-31G$^{**}$ basis set.[88] The relaxed radial dimensions of the different SiNTs and SiNWs are summarized in Table 1. Coordinates of the fully relaxed structures can be found in the supplementary material.

|  | Diameter [nm] | | | | | | | | |
|---|---|---|---|---|---|---|---|---|---|
|  | SiNW | | | SiNT – Outer | | | SiNT – Inner | | |
|  | LDA | PBE | HSE | LDA | PBE | HSE | LDA | PBE | HSE |
| 100 | 1.76 | 1.78 | 1.77 | 1.80 | 1.83 | 1.82 | 0.65 | 0.70 | 0.69 |
| 110 | 2.63 | 2.66 | 2.65 | 2.61 | 2.65 | 2.64 | 1.36 | 1.38 | 1.38 |
| 111 | 1.77 | 1.80 | 1.79 | 1.81 | 1.86 | 1.84 | 0.89 | 0.92 | 0.91 |
| 112 | 1.81 | 1.84 | 1.82 | 1.79 | 1.84 | 1.83 | 0.45 | 0.46 | 0.45 |

*Table 1: Average radial dimensions of the fully fluorinated SiNTs and SiNWs structures optimized using the LDA, PBE, and HSE exchange-correlation functional approximations and the 6-31G$^{**}$ basis set. The [110] dimensions have been obtained using only the 3-21G basis set due to the computational burden. Reported diameters are twice the average radius from the axis of the SiNW/SiNT to the external layer of silicon atoms.*

## **Stability**

We start by analyzing the relative structural stability of the different NWs and NTs shown in Fig. 1. As the SiNWs and SiNTs structures have different chemical compositions the cohesive energy per atom does not provide a suitable measure for the comparison of their relative stability. Therefore, we define the Gibbs free energy of formation ($\delta G$) for a SiNT and a SiNW as:[26,27,70,89]

$$\delta G(\chi_i) = E(\chi_i) - \sum \chi_i \mu_i \qquad (1)$$

where $E(\chi_i)$ ($i$=Si, H, F) is the cohesive energy per atom of a SiNW/T of a given composition, $\chi_i$ is the molar fraction of atom $i$ in the system with $\sum_i \chi_i = 1$, and $\mu_i$ is the chemical potential of element $i$. Here, we choose $\mu_H$ and $\mu_F$ as the binding energy per atom of the ground state of the H$_2$ and F$_2$ molecules, respectively, and $\mu_{Si}$ as the cohesive energy per atom of bulk silicon. This definition allows for a direct energetic comparison between SiNW/T with different chemical compositions, where negative values represent stable structures with respect to the constituents. It should be stressed that this treatment gives a qualitative measure of the relative stability while neglecting thermal and substrate effects and zero point energy corrections.[27]

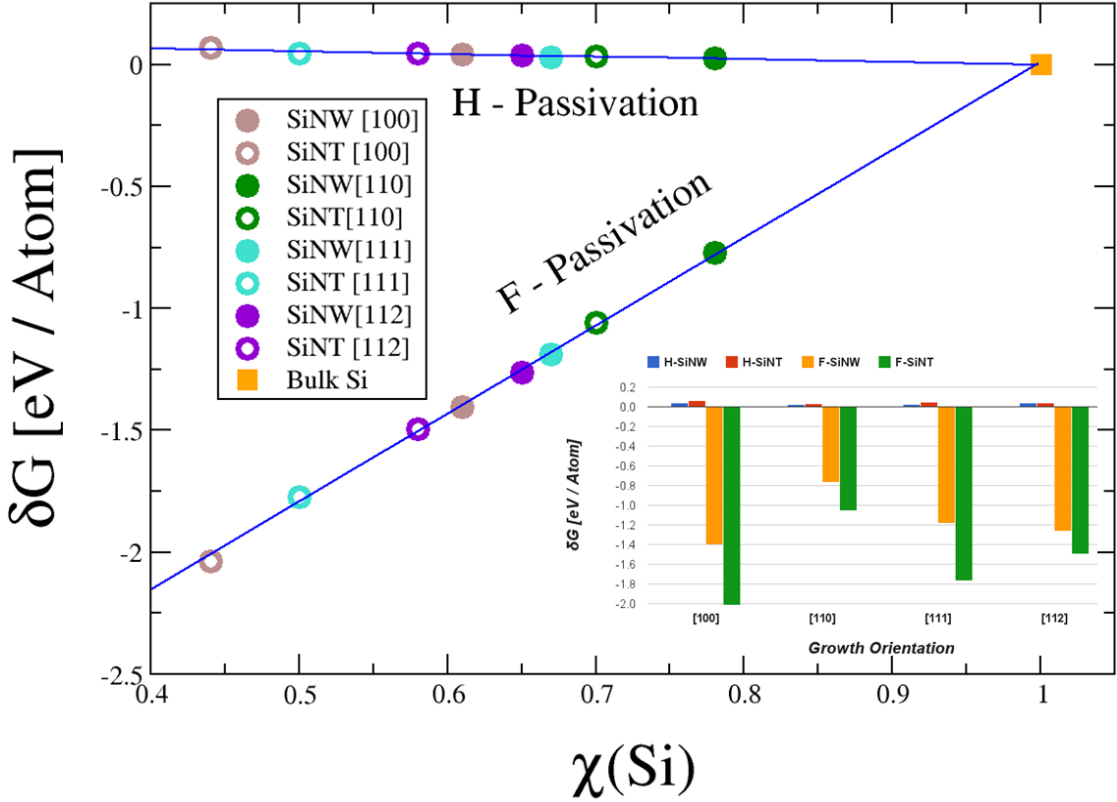

Figure 2: *δG as calculated using Eq. 1 at the HSE/6-31G\*\* level of theory for the different fully hydrogenated and fully fluorinated SiNWs and SiNTs considered. Inset: comparison of δG values between fully hydrogenated and fully fluorinated SiNWs and SiNTs models with similar growth orientation. To reduce computational burden geometry optimization of the fluorinated [110] SiNW and SiNT has been performed at the LDA/3-21G and HSE/3-21G level of theory, respectively. Single-point calculations performed at the HSE/6-31G\*\* level of theory using the obtained geometries of both structures were used to evaluate δG. We estimate that this procedure introduces errors of less than 40% in the calculated values (see supplementary material).*

In Fig. 2 we plot δG as calculated using Eq. 1 for the fully fluorinated SiNWs and SiNTs considered. For comparison we include the results obtained in Ref. [70] for the fully hydrogenated systems. Similar to the case of the latter, we identify a linear relation between the molar fraction of the silicon atoms and the relative stability of the fully fluorinated SiNWs/SiNTs considered. Nevertheless, while the stability of the various hydrogenated systems was found to decrease with decreasing silicon molar fraction the fluorinated structures present a completely opposite picture both qualitatively and quantitatively. When comparing the hydrogenated structures to their fluorinated counterparts a considerable increase in stability is obtained upon fluorine decoration. The stability is further enhanced as the molar fraction of the silicon content decreases. Furthermore, unlike the case of the hydrogen passivated structures it is found that the fluorinated SiNTs considered are more stable than their SiNWs counterparts (see inset of Fig. 2).

These observations can be rationalized by considering the dissociation energies of Si-F, Si-H and Si-Si bonds, which were measured to be ~160, ~90, (in (fluoro)silanes) and 54 kcal/mol (in bulk form), respectively.[90] As can be seen, the Si-F bond is about 1.8 times stronger than Si-H bond. In fact, the Si-F bond is among the strongest single bonds known thus explaining the increased stability upon replacement of surface hydrogens with fluorine atoms. Furthermore, since we are considering relatively narrow NTs, the effect of extraction of the inner silicon core on the stability of the system is fully compensated by the Si-F bonds formation on the inner surface thus resulting is systems which are even more stable than the corresponding NWs.

Next, we examine the relative stability of the [100] and [112] SiNWs and SiNTs with mixed hydrogenation and fluorination decoration schemes. Fig. 3 shows the calculated δG values for these systems as a function of fluorine surface coverage where 0% represent the fully hydrogenated systems and 100% stands for the fully fluorinated systems. As may be expected from the analysis presented above, the stability increases with increasing fluorine coverage. Interestingly, already at ~25% fluorine coverage the SiNTs structures become more stable than their corresponding SiNWs. For each mixed decorated system three random decoration schemes per given fluorine coverage were considered. We find that the differences between their relative stabilities were of the order of 0.01 eV/atom and thus are too small to be noticed on the diagrams presented in Fig. 3.

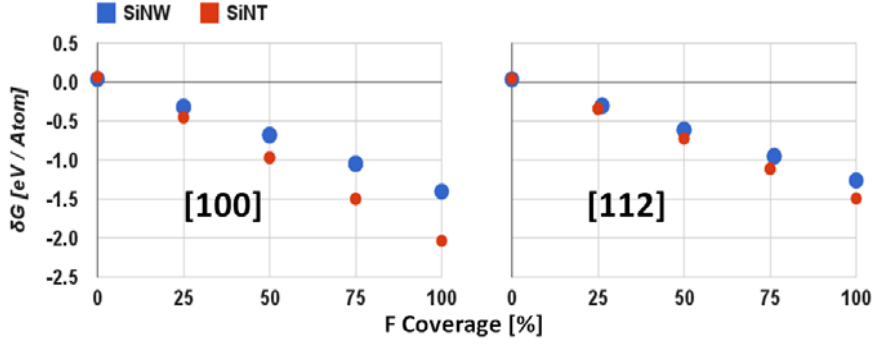

*Fig. 3: δG as calculated using Eq. 1 for the [100] (left panel) and [112] (right panel) SiNWs and SiNTs with varying fluorine and hydrogen coverage. For the [112] SiNT with 50% fluorine coverage only two (instead of three) random decoration schemes have been calculated.*

**Electronic properties**

We now turn to study the electronic properties of the various systems considered. We start by analyzing the effects of crystallographic orientation and surface passivation scheme on the bandgap of the SiNWs and their corresponding SiNTs studied. Fig. 4 presents a detailed description of these results where the bandgap is plotted against the molar fraction of silicon atoms within each system. As can be seen, similar to the case of the fully hydrogenated systems[70] (large orange circles) the bandgap of the fully fluorinated SiNWs and SiNTs (dark purple circles) decreases monotonously with increasing silicon molar fraction gradually approaching the bandgap value of bulk silicon. For low silicon molar fractions the bandgap of the fully fluorinated systems is narrower than that of the fully hydrogenated structures by up to 0.79 eV. As the silicon molar fraction increases the difference in bandgap of the fluorinated and hydrogenated systems reduces such that for molar fractions exceeding 0.65 the bandgaps are identical within 0.02 eV. Furthermore, as may be expected, all fully fluorinated SiNTs considered possess a wider bandgap than their corresponding SiNWs. This can be attributed to quantum confinement effects. From the inset of Fig. 4 it can be clearly deduced that the effect of the appearance of the inner cavity on the bandgap of the fully fluorinated systems is smaller than that previously obtained for the hydrogenated systems. The transition from [111], [100], [112] and [110] F-SiNWs to F-SiNTs results in a bandgap increase of 24, 41, 11 and 11%, respectively. This is to be compared with the hydrogenated systems values of 68, 61, 26 and 17%, respectively.

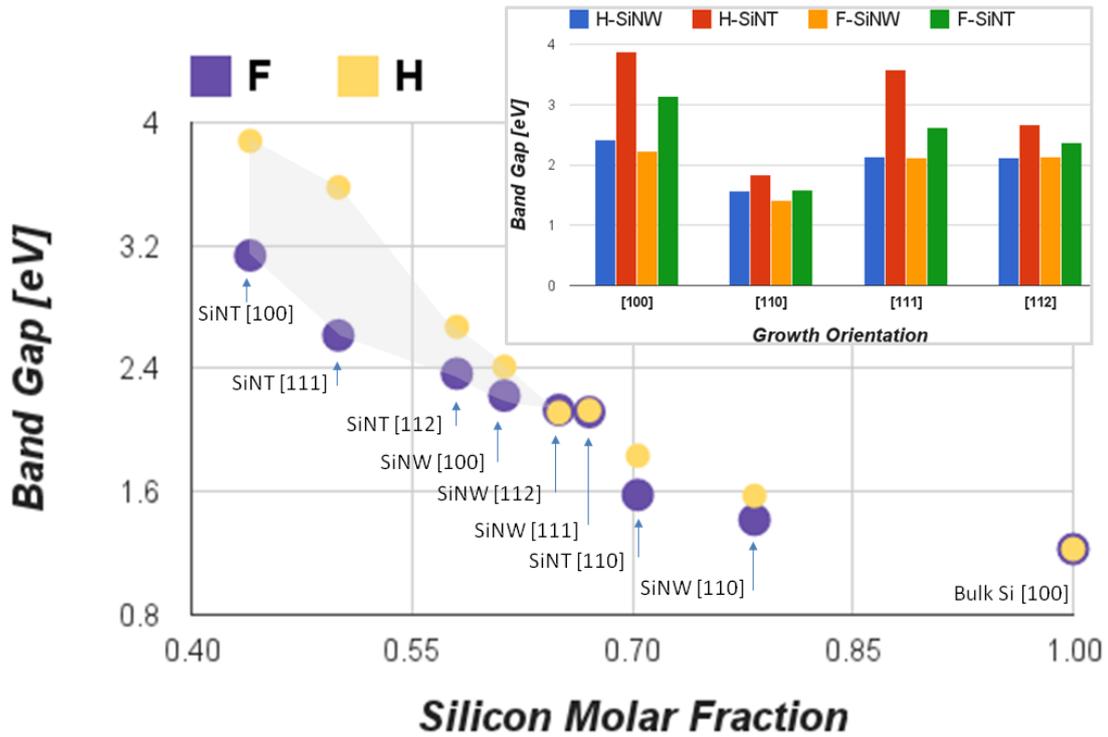

Fig. 4: *Bandgap vs. Silicon molar fraction for the fluorinated SiNWs and SiNTs studied (blue circles). Results for the fully hydrogenated system are taken from Ref.[70] and are presented for comparison purposes (orange circles). The transparent gray area indicates the region spanned between the bandgap values of the fully hydrogenated and fluorinated systems. Inset: Comparison between bandgaps of the fully hydrogenated and fluorinated SiNWs and SiNTs at different growth orientations. To reduce computational burden geometry optimization of the fluorinated [110] SiNW and SiNT has been performed at the LDA/3-21G and HSE/3-21G level of theory, respectively. Single-point calculations performed at the HSE/6-31G\*\* level of theory using the obtained geometries of both structures were used to evaluate the bandgaps. We estimate that this procedure introduces errors of less than 8% in the calculated values (see supplementary material).*

Upon careful inspection of the results appearing in Fig. 4 it can be speculated that by varying the relative surface coverage of hydrogen and fluorine atoms it should be possible to tailor the bandgap of SiNWs and SiNTs having a silicon molar fraction lower than 0.65. In the diagram, we highlight the area spanning the region between the fully hydrogenated and fully fluorinated systems bandgaps indicating the expected bandgaps that can be achieved by manipulating the surface fluorination scheme. To further examine this issue, we study in Fig. 5 the influence of fluorine coverage on the bandgap of the [100] and the [112] systems. As mentioned above, from quantum confinement considerations the bandgaps of the SiNTs are consistently larger than those of their corresponding SiNWs. As the fluorine coverage is increased the bandgaps of the SiNTs decrease monotonously approaching that of the fully fluorinated systems. On the contrary, the bandgaps of the SiNWs are much less sensitive to the fluorine coverage such that the values calculated for the fully fluorinated and fully hydrogenated systems differ by less than 0.2 eV as is also indicated in Fig. 4.

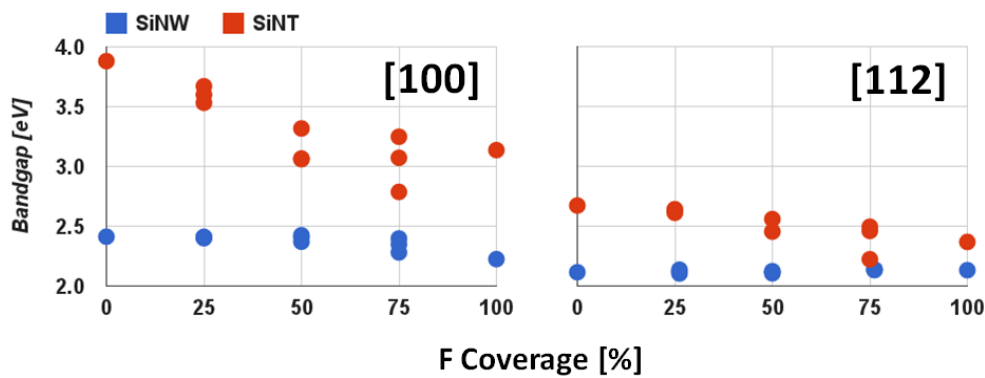

Fig. 5: *Bandgaps of the [100] (left panel) and [112] (right panel) SiNWs (blue marks) and SiNTs (red marks) as a function of fluorine coverage. For the [112] SiNT with 50% fluorine coverage only two (instead of three) random decoration schemes have been calculated.*

Interestingly, for the mixed decoration schemes at given fluorine coverage a strong effect of the exact decoration scheme on the bandgap of the system is obtained. We find that for the 50% and 75% fluorine covered [100] and for the 75% covered [112] SiNTs the bandgap of part of the randomly chosen decoration arrangements is smaller than that of the corresponding fully fluorinated system. Here, we should note that our calculations are of periodic nature. In realistic experimental conditions different sections along the system will have varying passivation schemes and thus we expect the bandgap of the system at a given fluorine coverage to be averaged.

The results presented in Fig. 5 clearly indicate that surface chemistry may be used to tailor the electronic properties of narrow quasi-one-dimensional silicon nanostructures. By careful surface functionalization we predict that it should be possible to tune the bandgap of such systems to range between insulating to wide-bandgap semi-conducting even without manipulating the structure of the inner silicon core.

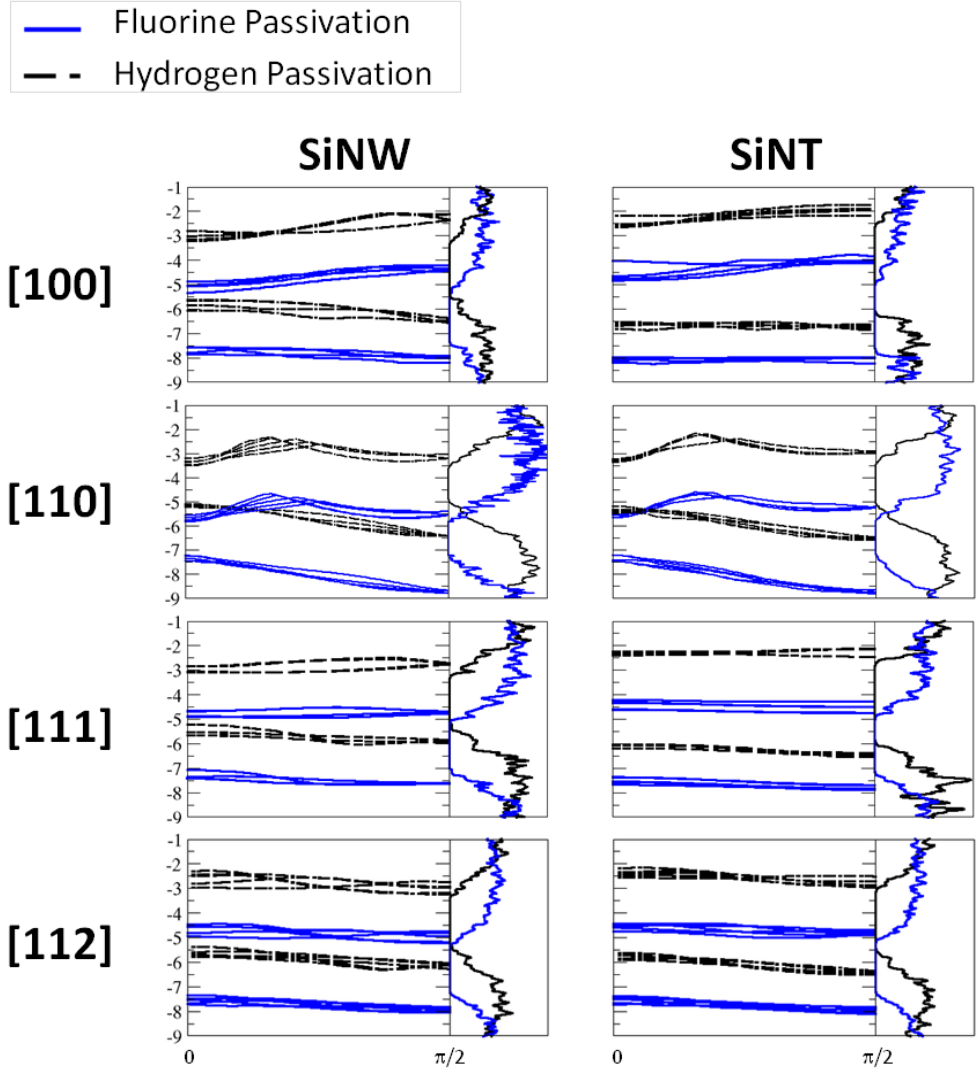

*Fig. 6: Band structure and density of states of the fully fluorine decorated SiNWs and SiNTs studied. Results for the fully hydrogenated systems are presented for comparison. For clarity, only five occupied and five virtual bands are presented.*

To gain further insight regarding the influence of fluorination on the electronic properties of SiNWs and SiNTs we present in Fig. 6 the band structures and the corresponding densities of states of the fully fluorinated systems and compare them to the results obtained for their fully hydrogenated counterparts. Apart from the fluorination induced bandgap reduction discussed above a pronounced downshift of both the occupied and unoccupied bands is observed for both the SiNWs and SiNTs regardless of the crystallographic orientation. For the systems considered the valence band maxima (VBM) downshifted by up to 2.16eV and the conduction band minima (CBM) downshifted by up to 2.32 eV. Interestingly, when comparing the VBMs and CBMs of the fully hydrogenated and fluorinated systems it is found that their dispersion relation is hardly affected upon fluorination and that they are practically rigidly downshifted. The effect of bandgap reduction thus results from a different rigid downshift that the VBM and CBM of each system experience. For the same reasons the nature of the bandgap of all but the [111] SiNW systems studied is not affected by fluorination such that both the fully hydrogenated and fluorinated [100] systems present a direct bandgap whereas the [112] systems and the [111] SiNT present an indirect gap. The [111] SiNW exhibits a transition from direct to indirect bandgap upon fluorination. For all the fluorinated systems studied we find that the wires and the tubes exhibit the same bandgap character.

**Summary**

In this paper, we presented a theoretical study of the structural and electronic properties of fluorine decorated narrow sp$^3$ type SiNTs and SiNWs bearing a wall thickness of a few atomic layers. Eight SiNT and SiNW models with periodic axes along the [100], [110], [111], and [112] crystallographic orientations were considered. The energetic stability and electronic properties of the various fluorinated systems considered were compared to those of their hydrogen passivated SiNTs and SiNWs counterparts. Unlike the fully hydrogenated systems, the relative stability of the fully fluorinated structures was found to increase linearly with decreasing silicon molar fraction. Furthermore, the fully fluorinated systems were found to be consistently more stable than their hydrogenated counterparts. For mixed hydrogenation and fluorination decoration schemes of the [100] and [112] SiNTs and SiNWs the relative stability was found to vary linearly with the fluorine surface coverage. Interestingly, for fluorine surface coverage exceeding 25% the tubular structures were found to be more stable than their wire counterparts.

Similar to the case of the fully hydrogenated systems the bandgaps of the fully fluorinated structures were found decrease monotonously with increasing silicon molar fraction. For silicon molar fractions lower than 0.5 the bandgap value for fully fluorinated systems is smaller than that of the fully hydrogenated structures by up to 0.79 eV. These differences reduce as the silicon molar fraction increases. As may be expected, the fully fluorinated SiNTs bear larger bandgaps than their corresponding SiNWs. Nevertheless, the differences between their bandgaps are lower than those obtained for the hydrogenated systems. The mixed hydrogenated and fluorinated systems usually present bandgaps that reside within the range spanned by the fully hydrogenated and fully fluorinated structures. Some exceptions occur where the bandgaps of the mixed decorated systems become lower than the bandgaps of their fully fluorinated counterparts. In general the bandgaps of the SiNTs studied are more sensitive to the fluorine surface coverage than those of the corresponding SiNWs. Upon examination of the band-structure of the various systems, the reduction of the bandgap upon fluorination was found to result from a different and (almost) rigid downshift of the valence and conduction bands in the vicinity of the gap.

Our results indicate that surface functionalization may serve to control the structural stability of narrow quasi-one-dimensional silicon nanostructures and open the way towards chemical tailoring of their electronic properties. We predict that by careful design of the chemical surface decoration scheme it should be possible to tune the bandgap of such systems to range between insulating to wide-bandgap semi-conducting even without manipulating the structure of the inner silicon core.

## Acknowledgments

The authors would like to thank Professor Fernando Patolsky for helpful and insightful discussions. This work was supported by the Israel Science Foundation under Grant No. 1313/08, the Center for Nanoscience and Nanotechnology at Tel Aviv University, and the Lise Meitner-Minerva Center for Computational Quantum Chemistry.